\documentclass[11pt]{article}
\pdfoutput=1

\usepackage{amsmath,amsfonts,enumerate,array}
\usepackage{graphicx}
\usepackage[usenames]{color}
\usepackage[english]{babel}
\usepackage{fancybox}
\usepackage{chngpage} % allows for temporary adjustment of side margins

\usepackage{physymb}

\usepackage{calligra}
\DeclareMathAlphabet{\mathcalligra}{T1}{calligra}{m}{n}
\DeclareFontShape{T1}{calligra}{m}{n}{<->s*[2.2]callig15}{}
\newcommand{\scripty}[1]{\ensuremath{\mathcalligra{#1}}\;\!}

\newcommand{\captionfonts}{\small}

\makeatletter  % Allow the use of @ in command names
\long\def\@makecaption#1#2{%
  \vskip\abovecaptionskip
  \sbox\@tempboxa{{\captionfonts #1: #2}}%
 \ifdim \wd\@tempboxa >\hsize
    {\captionfonts #1: #2\par}
  \else
    \hbox to\hsize{\hfil\box\@tempboxa\hfil}%
  \fi
  \vskip\belowcaptionskip}
\makeatother   % Cancel the effect of \makeatletter

\usepackage{mciteplus}

\usepackage[colorlinks=true,      linkcolor=blue,      urlcolor=blue,      filecolor=blue,      citecolor=blue,
      pdfstartview=FitH,     pdfpagemode=UseNone,      bookmarksopen=true]{hyperref}
\usepackage[all]{hypcap}     %should be loaded AFTER hyperref, which should otherwise be loaded last.

\begin{document}

\numberwithin{equation}{section}

%%%%%%%%%%%%%%%%%%%%%%%%%%%%%%%%%%%%%%%%%%%%%%%%%%%%%%%%%%%%
%                       DEFINITIONS

\mathchardef\mhyphen="2D

%%%%%%%%%%%%%%%%%%%%%%%%%%%%%%%%%%%%%%%%%%%%%%%%%%%%%%%%%%%%
%                        Commands

\newcommand{\be}{\begin{equation}}
\newcommand{\ee}{\end{equation}}
\newcommand{\bea}{\begin{eqnarray}\displaystyle}
\newcommand{\eea}{\end{eqnarray}}
\newcommand{\nnm}{\nonumber}
\newcommand{\nn}{\nonumber}

\def\eq#1{(\ref{#1})}

%%%%%%%%%%%%%%%%%%%%%%%%%%%%%%%%%%%%%%%%%%%%%%%%%%%%%%%%%%%%%
%                        Greek letters

\def\a{\alpha}  \def\b{\beta}   \def\c{\chi}    
\def\g{\gamma}  \def\G{\Gamma}  \def\e{\epsilon}  
\def\vep{\varepsilon}   \def\tvep{\widetilde{\varepsilon}}
\def\f{\phi}    \def\F{\Phi}  \def\fb{{\ov \phi}}
\def\vf{\varphi}  \def\m{\mu}  \def\mub{\ov \mu}
\def\n{\nu}  \def\nub{\ov \nu}  \def\o{\omega}
\def\O{\Omega}  \def\r{\rho}  \def\k{\kappa}
\def\kab{\ov \kappa}  \def\s{\sigma}
\def\t{\tau}  \def\th{\theta}  \def\sb{\ov\sigma}  \def\S{\Sigma}
\def\l{\lambda}  \def\L{\Lambda}  \def\p{\psi}

%%%%%%%%%%%%%%%%%%%%%%%%%%%%%%%%%%%%%%%%%%%%%%%%%%%%%%%%%%%%%
%              Calligraphic & Blackboard letters etc

\def\cA{{\cal A}} \def\cB{{\cal B}} \def\cC{{\cal C}}
\def\cD{{\cal D}} \def\cE{{\cal E}} \def\cF{{\cal F}}
\def\cG{{\cal G}} \def\cH{{\cal H}} \def\cI{{\cal I}}
\def\cJ{{\cal J}} \def\cK{{\cal K}} \def\cL{{\cal L}}
\def\cM{{\cal M}} \def\cN{{\cal N}} \def\cO{{\cal O}}
\def\cP{{\cal P}} \def\cQ{{\cal Q}} \def\cR{{\cal R}}
\def\cS{{\cal S}} \def\cT{{\cal T}} \def\cU{{\cal U}}
\def\cV{{\cal V}} \def\cW{{\cal W}} \def\cX{{\cal X}}
\def\cY{{\cal Y}} \def\cZ{{\cal Z}}

\def\mC{\mathbb{C}} \def\mP{\mathbb{P}}  
\def\mR{\mathbb{R}} \def\mZ{\mathbb{Z}} 
\def\mT{\mathbb{T}} \def\mN{\mathbb{N}}
\def\mH{\mathbb{H}} \def\mX{\mathbb{X}}

\newcommand{\rmd}{\mathrm{d}}
\newcommand{\rmx}{\mathrm{x}}

\def\one{{\hbox{\kern+.5mm 1\kern-.8mm l}}}
\def\zero{{\hbox{0\kern-1.5mm 0}}}

%%%%%%%%%%%%%%%%%%%%%%%%%%%%%%%%%%%%%%%%%%%%%%%%%%%%%%%%%%%%%%
%                                 QM               

\newcommand{\bra}[1]{{\langle {#1} |\,}}
\newcommand{\ket}[1]{{\,| {#1} \rangle}}
\newcommand{\braket}[2]{\ensuremath{\langle #1 | #2 \rangle}}
\newcommand{\Braket}[2]{\ensuremath{\langle\, #1 \,|\, #2 \,\rangle}}
\renewcommand{\norm}[1]{\ensuremath{\left\| #1 \right\|}}
\def\corr#1{\left\langle \, #1 \, \right\rangle}
\def\vac{|0\rangle}

%%%%%%%%%%%%%%%%%%%%%%%%%%%%%%%%%%%%%%%%%%%%%%%%%%%%%%%%%%%%%
%                         General

\def\d{ \partial } 
\def\zb{{\bar z}}

\newcommand{\sq}{\square}
\newcommand{\IP}[2]{\ensuremath{\langle #1 , #2 \rangle}}    %inner product

\newcommand{\floor}[1]{\left\lfloor #1 \right\rfloor}
\newcommand{\ceil}[1]{\left\lceil #1 \right\rceil}

\newcommand{\T}[3]{\ensuremath{ #1{}^{#2}_{\phantom{#2} \! #3}}}		%general tensor with upper indices displayed first 

\def\ha{\frac{1}{2}}
\def\tha{\tfrac{1}{2}}
\def\wt{\widetilde}
\def\ra{\rangle}
\def\la{\langle}

\def\ov{\overline}
\def\Slash{\, / \! \! \! \!}
\def\dslash{\partial\!\!\!/} 
\def\Dslash{D\!\!\!\!/\,\,}
\def\fslash#1{\slash\!\!\!#1}
\def\Fslash#1{\slash\!\!\!\!#1}

\def\del{\partial}
\def\delb{\bar\partial}
\newcommand{\ex}[1]{{\rm e}^{#1}} 
\def\ii{{i}}

\newcommand{\ap}{\ensuremath{\alpha'}}

\newcommand{\bean}{\begin{eqnarray*}}
\newcommand{\eean}{\end{eqnarray*}}

%%%%%%%%%%%%%%%%%%%%%%%%%%%%%%%%%%%%%%%%%%%%%%%%%%%%%%%%%%%%%%%%%%%%%%%%%%%%%

\def\q{\quad}

\def\bn{B_\circ}

\let\a=\alpha \let\b=\beta \let\g=\gamma 
\let\e=\epsilon
\let\c=\chi \let\th=\theta  \let\k=\kappa
\let\l=\lambda \let\m=\mu \let\n=\nu \let\x=\xi \let\r=\rho

\let\s=\sigma 
\let\t=\tilde

\def\r{\rightarrow}
\def\Ri{\Rightarrow}

\def\nn{\nonumber\\}
\def\b{\bigskip}
\def\b{\vspace{3mm}}

\let\p=\partial

\makeatletter
\def\blfootnote{\xdef\@thefnmark{}\@footnotetext}  % for blank footnote
\makeatother

%%%%%%%%%%%%%%5%%%%%%%%%%%% Reference Shortcuts %%%%%%%%%%%%%%%%%%%%%%%%%%%%%%%%%%%%

\newcommand{\spectral}{Schwimmer:1986mf}

\newcommand{\sv}{Strominger:1996sh}
\newcommand{\cvet}{Cvetic:1996xz,*Cvetic:1997uw}
\newcommand{\mm}{Balasubramanian:2000rt,*Maldacena:2000dr}
\newcommand{\MM}{Balasubramanian:2000rt,*Maldacena:2000dr}

\newcommand{\lmone}{Lunin:2001ew}
\newcommand{\lmtwo}{Lunin:2001pw}
\newcommand{\lmfour}{Lunin:2001jy}
\newcommand{\lmRot}{Lunin:2001fv}
\newcommand{\lmAdS}{Lunin:2001jy}
\newcommand{\lmm}{Lunin:2002iz}

\newcommand{\GMR}{Gutowski:2003rg}
\newcommand{\CarMcCon}{Cariglia:2004wu}

\newcommand{\lunin}{Lunin:2004uu}
\newcommand{\gmsone}{Giusto:2004id}
\newcommand{\gmstwo}{Giusto:2004ip}

\newcommand{\bw}{Bena:2004de}
\newcommand{\fuzz}{Mathur:2005zp,*Bena:2007kg,*Skenderis:2008qn,*Mathur:2012zp}
\newcommand{\kst}{Kanitscheider:2007wq}
\newcommand{\cern}{Mathur:2009hf}

\newcommand{\mtone}{Mathur:2011gz}
\newcommand{\mttwo}{Mathur:2012tj}
\newcommand{\lmt}{Lunin:2012gp}
\newcommand{\glmt}{Giusto:2012yz}

\newcommand{\orbifoldrefs}{Arutyunov:1997gt,*Arutyunov:1997gi,*deBoer:1998ip,*Dijkgraaf:1998gf,
*Seiberg:1999xz,*Larsen:1999uk,*David:1999zb,*Jevicki:1998bm}

%%%%%%%%%%%%%%%%%%%%%%%%%%%%%%%%%%%%%%%%%%%%%%%%%%%%%%%%%%%%%%%%%%%%%%%%%%%%%%%%%%%%%%%%%%%%%%

\begin{flushright}
%OHSTPY-HEP-T-03-012\\
\end{flushright}
\vspace{20mm}
\begin{center}
{\hbox {\LARGE  {Oscillating supertubes %} 
and neutral rotating black hole microstates}}}
\vspace{2cm}

{\bf Samir D. Mathur ~and~ David Turton} \\

\vspace{15mm}
Department of Physics,\\ The Ohio State University,\\ Columbus,
OH 43210, USA

\vspace {5mm}
mathur.16@osu.edu,~~turton.7@osu.edu
\vspace{4mm}

\end{center}

\vspace{1.5cm}

\begin{abstract}

The construction of neutral black hole microstates is an important problem, with implications for the information paradox.
In this paper we conjecture a construction of non-supersymmetric supergravity solutions describing D-brane configurations which carry mass and angular momentum, but no other conserved charges. 
We first study a classical string solution which locally carries dipole winding and momentum charges in two compact directions, but globally carries no net winding or momentum charge. 
We investigate its backreaction in the D1-D5 duality frame, where this object becomes a supertube which locally carries oscillating dipole D1-D5 and NS1-NS5 charges, and again carries no net charge.
In the limit of an infinite straight supertube, we find an exact supergravity solution describing this object.
We conjecture that a similar construction may be carried out based on a class of two-charge non-supersymmetric D1-D5 solutions. 
These results are a step towards demonstrating how neutral black hole microstates may be constructed in string theory.

\end{abstract}

\thispagestyle{empty}

\newpage

\baselineskip=15pt
\parskip=3pt

\section{Introduction}

The black hole information paradox has been a long-standing problem in physics~\cite{Hawking:1974sw,*Hawking:1976ra}. In recent years we have seen how the paradox may be resolved in string theory: the gravitational fields sourced by individual microstates have been found to differ from the traditional geometry with horizon (for reviews, see~\cite{Mathur:2005zp,*Bena:2007kg,*Skenderis:2008qn,*Chowdhury:2010ct,*Mathur:2012zp}). For the microstates which have been constructed,  in each known example it has been found that the microstate sources a horizonless gravitational solution. While there has been considerable work on supersymmetric microstate solutions\footnote{For more recent work, see e.g.~\cite{Bena:2011uw,*Bena:2011dd,*Gibbons:2013tqa,
Giusto:2011fy,*Giusto:2012gt,*Giusto:2012jx,*Giusto:2013rxa,
\mtone,*\mttwo,*\lmt,*\glmt}.}, the number of non-supersymmetric solutions known to date is small; see e.g.~\cite{Jejjala:2005yu,Giusto:2007tt,*AlAlawi:2009qe,Bobev:2011kk,*Vasilakis:2011ki,*Bena:2011ca,*Bena:2011fc,*Bena:2012zi,*Bena:2013gma}.

In considering non-BPS configurations, it is natural to seek a construction of gravitational solutions describing microstates of extremal, non-BPS black holes such as the extremal Kerr~\cite{Kerr:1963ud} and Myers-Perry~\cite{Myers:1986un} solutions. There is strong observational evidence of the existence of near-extremal black holes in our galaxy~\cite{McClintock:2006xd,*Gou:2013dna}. There has also been much interest in a recent conjecture that quantum gravity in the near-horizon region of an extremal Kerr black hole may be described by a chiral 2D CFT~\cite{Guica:2008mu}.

In this paper we describe progress on constructing new classes of non-supersymmetric D-brane configurations, which carry mass and angular momentum but no other conserved charges. Our starting point is a classical fundamental string solution which is a generalization of the solution studied in~\cite{Lunin:2001fv}. The configuration has the structure of a ring which locally carries dipole winding and momentum charges in two compact directions. The dipole charges rotate around the ring in such a way that globally the solution carries no net winding or momentum. 
This is achieved by taking the string to make a large circle in the covering space of a compact torus $T^2$. The string makes a closed orbit in the covering space, so has no net winding or momentum charge.

By considering a series of S and T dualities, we infer the existence of an `oscillating supertube': an object which locally carries  a combination of different dipole charges which oscillate as a null wave along the object. In the example we study, the charges oscillate between D1-D5 and
NS1-NS5. We argue that the configuration corresponding to the above fundamental string solution has the structure of a ring which has local oscillating D1-D5 and NS1-NS5 dipole charges, but which globally carries only mass and angular momentum.

By taking a near-ring limit, in which the configuration becomes an infinite straight line, we obtain an exact solution to IIB supergravity. The solution locally resembles a BPS supertube. We interpret the above as evidence that such an object exists in string theory, and for the existence of a solution for the ring configuration. 

We then conjecture that a similar construction may be carried out in the more general setting of the JMaRT family of non-supersymmetric microstate solutions~\cite{Jejjala:2005yu}. 
We consider the range of parameters of zero momentum charge, small D1 and D5 charges compared to the ADM mass, and small compact directions.
In this limit, the JMaRT solutions approach the 5D singly-spinning extremal Myers-Perry black hole (which has a zero-size horizon). 
We observe that the solutions in this case resemble a thin tube carrying D1-D5 charge inside a Myers-Perry background.
This tube can be thought of as resolving the naked ring singularity of the corresponding over-rotating Myers-Perry solution.
This motivates our conjecture that a similar construction may yield another new charge-neutral solution.

The circular oscillating supertube configurations are non-BPS, and so may be expected to decay. The configurations we study contain time-dependent gauge fields, which may be expected to be a source of decay\footnote{We thank Simon Ross and Nick Warner for raising the issue of  radiation from such solutions.}. A detailed analysis of the decay modes of these objects is beyond the scope of this paper, however we will make a simple estimate of the radiation from the circular ring configuration. It remains an exciting possibility that oscillating supertubes may be used in constructing microstates of neutral black holes such as extremal or near-extremal Kerr or Myers-Perry black holes, and possibly even Schwarzschild black holes.

Our work builds on classical string solutions studied previously in the literature. The classical string solution of interest to us is closely related to solutions studied in~\cite{Frolov:2003qc,Chialva:2003hg,*Chialva:2004xm,BlancoPillado:2007iz}. In particular,~\cite{BlancoPillado:2007iz} proposed a correspondence between a set of such configurations and black ring solutions with dipoles of the NS-NS two form potential~\cite{Emparan:2004wy}.

The plan of this paper is as follows. In Section 2 we present the classical string solution and analyze its properties. In Section 3 we consider the duality chain to the D1-D5 duality frame, and present our new supergravity solution describing an oscillating supertube. In Section 4 we conjecture a possible extension of these results to the two-charge JMaRT class of backgrounds. In Section 5 we discuss our results.

%\newpage

\section{The classical string solution}  \label{sec:string_soln}

In this section we consider the classical motion of a free string in flat space, without backreaction. We first recall the classical string equations. Then we write down the solution of interest to us; this solution will carry no overall winding or momentum charges.

\subsection{The equations for the string}

We briefly review the classical equations of motion for a string in flat space, which will enable us to introduce our notation and conventions.
String dynamics in flat space is described by the Nambu-Goto action
\be
S_{NG}=-T\sqrt{-\det[{\p X^\mu\over \p\chi^a}{\p X_\mu\over \p\chi^b}]}
\label{ng}
\ee
where
$
T={1\over 2\pi\alpha'}
$
is the tension of the string. Equivalently, one may use the Polyakov action
\be
S_P=-{T\over 2}\int d^2 \chi\sqrt{-g}{\p X^\mu\over \p\chi^a}{\p X_\mu\over \p \chi^b} g^{ab} \,.
\label{polyakov}
\ee
The variation of $g_{ab}$ gives
\be
{\p X^\mu\over \p\chi^a}{\p X_\mu\over \p \chi^b}-{1\over 2} g_{ab}{\p X^\mu\over \p\chi^c}{\p X_\mu\over \p \chi^d}g^{cd}=0
\ee
so $g_{ab}$ must be proportional to the induced metric. The $X^\mu$ equations give
\be
\p_a[\sqrt{-g}{\p X_\mu\over \p \chi^b} g^{ab}]=0
\ee
so that the $X^\mu$ are harmonic on the worldsheet. 

We choose coordinates $\chi^0\equiv {\tau},\, \chi^1\equiv {\sigma}$ on the world sheet so that $g_{ab}=e^{2\rho}\eta_{ab}$ for some $\rho$. Introducing
\be
\chi^+=\chi^0+\chi^1, ~~~\chi^-=\chi^0-\chi^1
\ee
we obtain
\be
g_{++}=0 \,, ~g_{--}=0 \,.
\label{conformal}
\ee
Since the induced metric is proportional to $g_{ab}$, this implies that
\be
{\p X^\mu\over \p\chi^+}{\p X_\mu\over \p \chi^+}=0 \,, ~~~{\p X^\mu\over \p\chi^-}{\p X_\mu\over \p \chi^-}=0 \,.
\label{condition}
\ee
To summarize, we solve the equations of motion if $X_\mu$ are harmonic functions
\be
X_\mu^{,a}{}_{a}=0
\label{xsol}
\ee
and they satisfy (\ref{condition}). The equations (\ref{xsol}) imply that the coordinates $X^\mu$ can be expanded as a left and a right moving part
\be
X^\mu=X^\mu_+(\chi^+)+X^\mu_-(\chi^-) \,.
\ee

\subsection{The classical string solution} \label{sec:stringsol}

We work in type IIB string theory with Minkowski spacetime compactified as
\be
M_{9, 1}~\r~ M_{4,1 }\times S^1\times T^4 \,.
\ee
We describe the noncompact part $M_{4,1}$ by coordinates $X^0=t, X^1, X^2, X^3, X^4$ and the $S^1$ by $y$ with  $0\le y\le 2\pi R$. For the $T^4$ we use coordinates $z^5, z^6, z^7, z^8$ of which we single out one, say $z^6\equiv z$ for our construction below. 
We take the  $T^4$ to be a square torus, with each coordinate having period $2\pi R_z$.

We can use the residual diffeomorphism symmetry to set the harmonic function $X^0$ to be
\be
X^0=2 A \tau = A(\chi^++\chi^-) \,.
\label{btau}
\ee 
The full solution is given by 
\bea
t&=& A ( \chi^+ + \chi^-)\nn
y&=& \frac{A}{k}\sin(k\,\chi^+)\nn
z&=& \frac{A}{k}\cos(k\,\chi^+)\nn
X^1&=&A\cos(\chi^-)\nn
X^2&=&A\sin(\chi^-) \,.
\label{ansatz}
\eea
The string describes a circle of radius $A$ in the noncompact $X^1, X^2$ plane, through a left-moving wave on the string. 
The string also oscillates in the compact coordinates $y, z$ through a {\it right}-moving vibration.  

One may check that the equations \eq{condition} and \eq{xsol} are satisfied. Each function is also periodic under $\sigma\r \sigma+2\pi$,  so the string is closed as it should be. We thus have a valid solution to the string equations. Similar solutions have previously been considered in~\cite{Frolov:2003qc,Chialva:2003hg,*Chialva:2004xm,BlancoPillado:2007iz}.

We now examine the properties of the solution. We take the $(y,z)$ torus to be small compared to $A/k$,
\be\label{eq:params}
R_y \sim R_z \ll \frac{A}{k} \,.
\ee
Thus the string will locally wind many times around the torus as it goes around the circle in the $(X^1,X^2)$ plane. 

\begin{figure}[t!]
\begin{center}
\includegraphics[width=12cm]{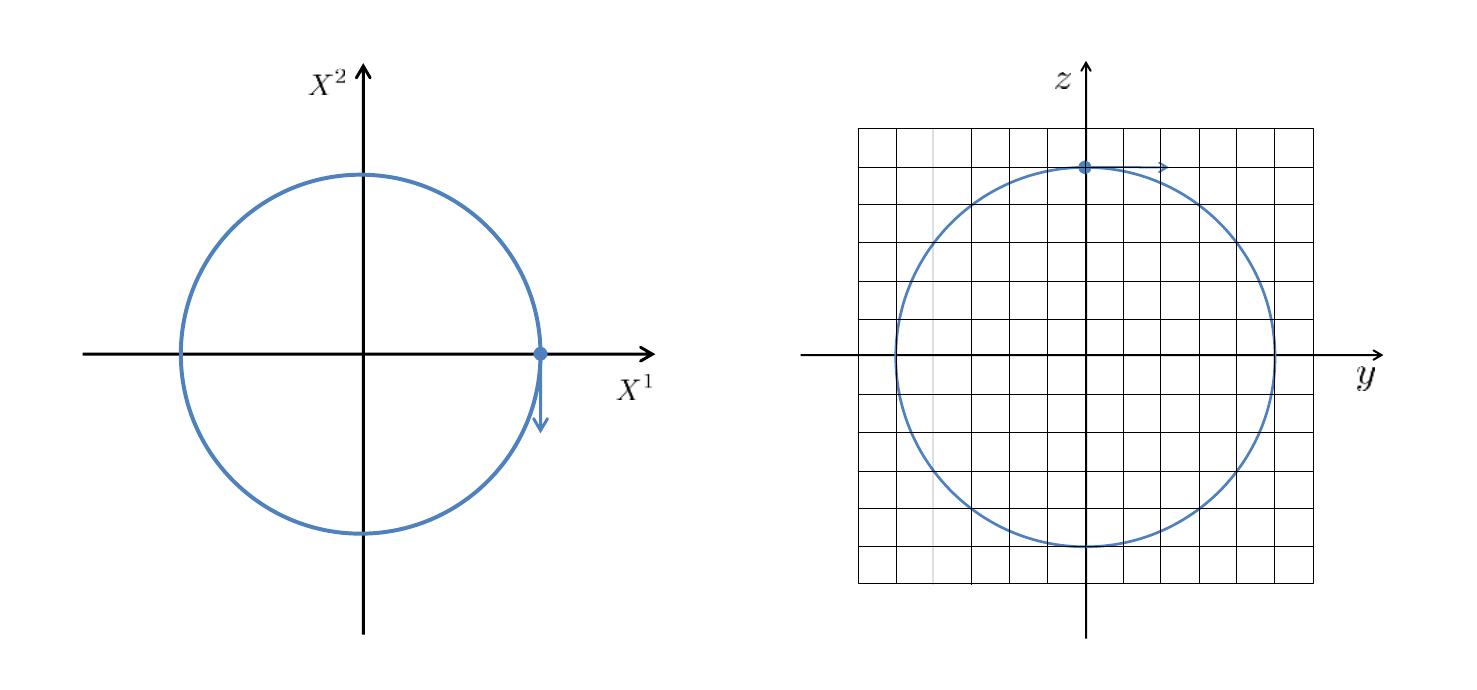}
\caption{String solution for $k=1$, at $\tau = 0$. The point $(\s,\tau)=(0,0)$ is marked with a blue dot and the direction of positive $\s$ is indicated with an arrow. At $(\s,\tau)=(0,0)$ the winding is predominantly in the positive $y$ direction. Since the string makes a closed loop in the covering space of the $(y,z)$ torus, there is no net winding charge.}
\label{fone}
\end{center}
\end{figure}

For ease of presentation, in the remainder of this section we shall analyze the case in which $k =1$.
This is depicted in Fig.\;\ref{fone}.

%
%

%\vspace{1cm}

\begin{figure}[t!]
\begin{center}
\vspace{5mm}
\includegraphics[width=11.5cm]{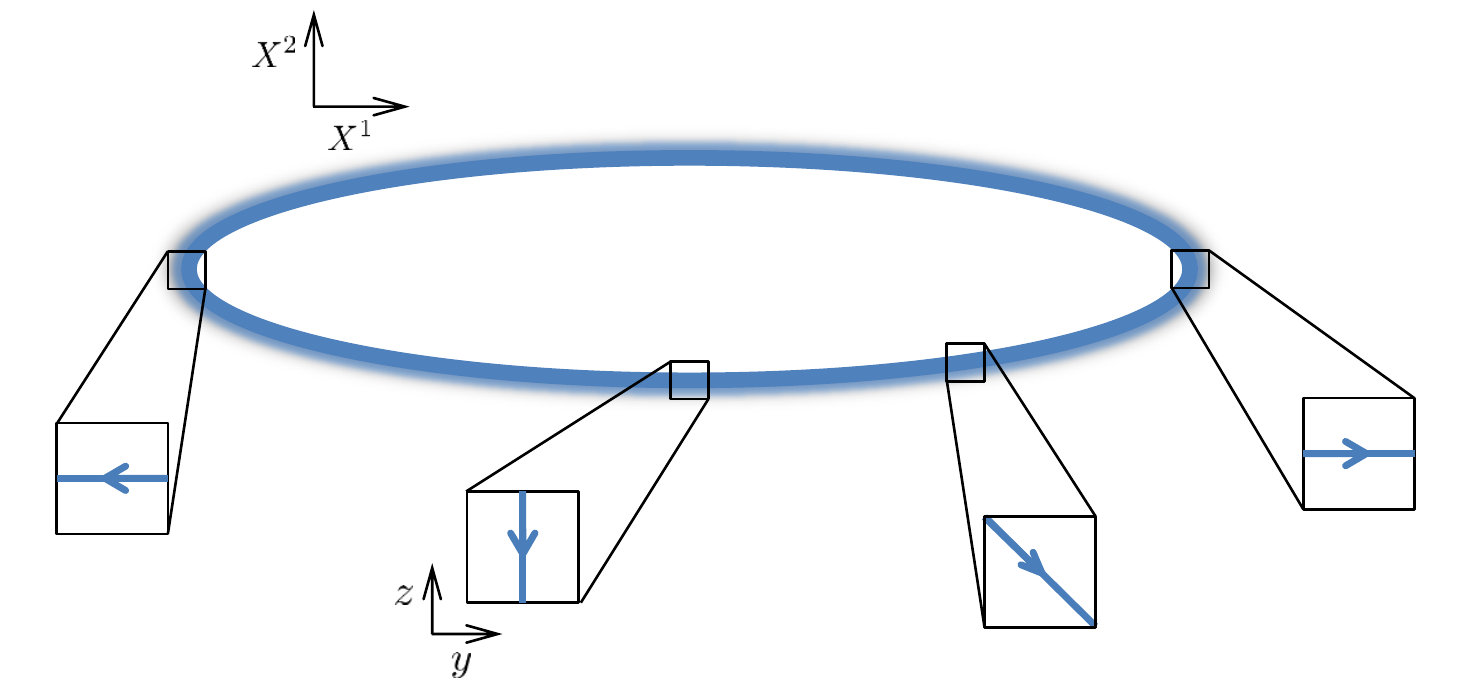}
\caption{A sketch of the configuration for $k=1$. The bold ring is a schematic depiction of the multiwound string. The boxes show in which direction the string winds in the $(y,z)$ plane at $\tau=0$. The string closes on itself after the full trajectory in the $(X^1,X^2)$ plane.}
\label{ftwo}
\end{center}
\end{figure}

At time $t=0$, we find as we circle the ring that
 \be
 \delta\chi^+=\delta\sigma, ~~~\delta\chi^-=-\delta\sigma \,.
 \ee
Let us examine a small segment of the circle around the point $(X^1,X^2)=(A, 0)$. We find
 \bea
 \delta y&\approx& A\delta\sigma\nn
 \delta z&\approx& 0\nn
\delta X^1&\approx&0\nn
\delta X^2&\approx& A\delta\sigma \,.
\eea
This is similar to the structure expected for a supertube where the string winds purely in the $y$ direction (i.e. with no excitation of the $z$ coordinate). As shown in Fig.\;\ref{ftwo}, the strands of the string wind partly along $y$ and partly along the $(X^1,X^2)$ circle; here the latter direction is along $X^2$. Thus the local charges of the supertube include (i) winding along $y$  and (ii) winding along $X^2$. The string strands also carry momentum. This momentum must be perpendicular to the direction of the strand, and we see from Fig.\;\ref{ftwo} that we get the local charges (iii) momentum along $y$ and (iv) momentum along $X^2$. (The latter is actually negative, as seen from the figure.)

Let us write $(X^1, X^2)=(A \cos\phi, A\sin\phi)$. Around a general point $\phi$ we find (at $t=0$)
 \bea
 \delta y&\approx& A\cos\phi\, \delta\sigma\nn
 \delta z&\approx& -A\sin\phi\, \delta \sigma\nn
\delta X^1&\approx& -A\sin\phi\, \delta \sigma\nn
\delta X^2&\approx& A\cos\phi\,  \delta \sigma
\eea
The compact directions have winding and momentum along $y$ as well as $z$. Note that the string does not have to close after each winding around the compact torus; it only has to close after a complete trip around the $(X^1,X^2)$ circle. Thus there is no constraint from the periodicity of the cycles on the torus. 

Breaking up the winding and momentum charges along different directions, we find that we have {\it six} local charges: windings along $\phi, y, z$ and momentum along $\phi, y, z$. 
When we perform dualities in the compact directions, the charges in these directions will change to other charges.

%\newpage
\section{The metric produced by the neutral oscillating supertube} \label{sec:D1D5ring}

The string profile (\ref{ansatz}) lies in a ring of radius $A$ in the $(X^1,X^2)$ plane. If we consider the backreaction of this string on the geometry, we will get a deformation of the metric in a tubular neighbourhood of this ring. Before we discuss this metric, we will perform some dualities to map the string source to other objects in the theory. The essential physics is, of course, unchanged under such duality, but with the string  we had a source singularity at the location of the string, while in a different duality frame the source is replaced by a KK monopole, and is hence the solution is a regular geometry.

\subsection{Dualizing to D1-D5}

We now consider the standard sequence of dualities which maps an NS1-P configuration to a D1-D5 bound state. We will use this duality map as a guide to motivate an ansatz for the fields describing the corresponding object in the new duality frame.

To get a handle on the problem, let us consider 
the neighborhood of 
the point $\chi^+ = 0$. Here we have
\be
y ~\simeq ~ A\chi^+  \,, \qquad z ~ \simeq ~  A \left( 1 - \ha (\chi^+)^2 \right) \,.
\ee
Let us also fix $\tau = 0$. At this instant we have $y ~\simeq ~ A\s$ so that (at $\s=0$) the string is locally winding in the $y$ direction only. A little away from $\s=0$ the string starts to wind slightly also in the $z$ direction (recall that the size of the $y$--$z$ torus is small compared to $A$). 

Let us therefore consider the sequence of dualities on a string carrying winding and momentum in the $y$ direction, and a smaller amount of winding and momentum in a $T^4$ direction. 
The penultimate step in this sequence of dualities involves T-duality along one torus direction. The starting string configuration also selects a particular torus direction. One obtains different results depending on whether these directions are taken to be the same or different. To illustrate this, we consider the two cases in which the string configuration winds around $z^6$ and $z^7$, and we consider T-duality along $z^6$ in each case.

We denote these starting configurations by
\be
\begin{array}{rlllllll}

~~~&~  \mathrm{NS}1_y & \mathrm{P}_y &   ~~~&~~~ (\mathrm{NS}1_6 & \mathrm{P}_6)	~~~&~~~ (\mathrm{NS}1_7 & \mathrm{P}_7)
\end{array}
\ee
where the brackets denote the two possibilities for the smaller charge.

We consider the following sequence of dualities:
\be
\begin{array}{r|lllllll}
   ~~~&~  \mathrm{NS}1_y & \mathrm{P}_y &   ~~~&~~~ (\mathrm{NS}1_6 & \mathrm{P}_6)	~~~&~~~ (\mathrm{NS}1_7 & \mathrm{P}_7)	
\cr 
\cr
\hline
\cr
S:   ~~~&~  \mathrm{D}1_y & \mathrm{P}_y &   ~~~&~~~ (\mathrm{D}1_6 & \mathrm{P}_6)	~~~&~~~ (\mathrm{D}1_7 & \mathrm{P}_7)	%~~& ~~~[\mathrm{IIB}] 
\cr\cr
T_{5678}:  ~~~&~  \mathrm{D}5_{5678y} & \mathrm{P}_y &   ~~~&~~~  (\mathrm{D}3_{578} & \mathrm{NS}1_{6}) ~~~&~~~  (\mathrm{D}3_{568} & \mathrm{NS}1_{7})	%~~& 
% ~~~[\mathrm{IIB}] 
 \cr\cr
S:   ~~~&~  \mathrm{NS}5_{5678y} & \mathrm{P}_y &   ~~~&~~~  (\mathrm{D}3_{578} & \mathrm{D}1_6)  ~~~&~~~  (\mathrm{D}3_{568} & \mathrm{D}1_7)%	~~& 
% ~~~[\mathrm{IIB}]  
\cr\cr
T_{y6}: ~~~&~  \mathrm{NS}5_{5678y} & \mathrm{NS}1_y ~~~& ~~~&~~~  (\mathrm{D}5_{5678y} & \mathrm{D}1_{y})
~~~&~~~  (\mathrm{D}3_{58y} & \mathrm{D}3_{67y})	
%~~& ~~~[\mathrm{IIB}] 
  \cr\cr
S:   ~~~&~   \mathrm{D}5_{5678y} & \mathrm{D}1_y & ~~~&~~~  (\mathrm{NS}5_{5678y} & \mathrm{NS}1_{y}) ~~~&~~~  (\mathrm{D}3_{58y} & \mathrm{D}3_{67y}) %~~&
%~~~[\mathrm{IIB}]  
\end{array}
\ee
Thus, in the case where the string is wound around $z^6$, we find that the resulting configuration has D1 and D5 charges along with small amounts of NS1 and NS5 charge. At generic points on the ring we will have all four charges locally, and at certain points the NS1 and NS5 charges will dominate over the D1 and D5 charges. 

In the case where the string is wound around $z^7$, the smaller local charge is (D$3_{58y}$~~D$3_{67y}$).
Lifting these charges to M-theory we obtain\footnote{We thank N.~Warner for a discussion on this point.}
\be
\begin{array}{r|lllllll}
   ~~~&~   \mathrm{D}5_{5678y} & \mathrm{D}1_y  & &
   ~~~&~~~  (\mathrm{D}3_{58y} & \mathrm{D}3_{67y}) ~~&
\quad ~~~[\mathrm{IIB}] \cr\cr 
T_{y78}:   ~~~&~   \mathrm{D}2_{56} & \mathrm{D}2_{78} & ~~~&~~~  
 ~~~&~~~  (\mathrm{D}2_{57} & \mathrm{D}2_{68}) ~~&
\quad ~~~[\mathrm{IIA}] \cr\cr 
\mathrm{lift}:   ~~~&~   \mathrm{M}2_{56} & \mathrm{M}2_{78} & ~~~&~~~  
 ~~~&~~~  (\mathrm{M}2_{57} & \mathrm{M}2_{68}) ~~&
\quad ~~~[\mathrm{M}] %\cr 
\end{array}
\ee
In this case the configuration has the local form of an M2-M2 supertube, and along the ring the wrappings of the M2 branes rotate in the torus directions.

In the following we will consider  the case where the string is wound around $z^6$, which is more convenient to analyze in the D1-D5 frame.
At opposite points on the string worldsheet ($\s$ versus $\s+\pi$), the winding is in the opposite direction around the compact directions, so there is no overall winding charge. Thus for the $z^6$ configuration,  the dual will have dipole D1, D5, NS1, NS5 charges but no overall net charge. The above considerations motivate the ansatz we will make in the next section, which will resemble a ``time-dependent S-duality" along the worldvolume of the supertube\footnote{For a similar structure in a different context, see \cite{Pilch:2000fu}.}.

\newpage
\subsection{The starting D1-D5 solution} \label{sec:MM_ring}

We work with type IIB string theory with the compactification
\be
M_{9, 1} ~~\r~~ M_{4, 1}\times S^1\times T^4 \,.
\ee
We wrap $n_1$ D1 branes on $S^1$, $n_5$ D5 branes on $S^1\times T^4$, and consider the bound states of these branes. 

We start by considering the geometry which is U-dual to a classical string configuration similar to that studied in Section \ref{sec:stringsol}, but which winds only around the $y$ direction and so carries both winding and momentum charge along this direction. After the dualities, these become D1 and D5 charges.

This geometry corresponds to the Ramond ground state obtained by spectral flow of the NS vacuum. It was first studied in~\cite{Balasubramanian:2000rt,*Maldacena:2000dr}
and takes the form:
\bea
ds^2&=& -\frac{1}{h} \left(dt + \frac{a\,Q}{f} \sin^2 \theta d\phi \right)^2
+ \frac{1}{h} \left(dy - \frac{a\,Q}{f} \cos^2 \theta d\psi \right)^2 \cr
&& {}+hf \left( \frac{dr^2}{r^2+a^2} + d\theta^2 \right) 
+h \Big[ r^2 \cos^2\theta d\psi^2 + (r^2+a^2) \sin^2\theta d\phi^2 \Big]
+
dz^i dz^i \,, \cr
C^{(2)}_{ty}&=&-{Q\over Q+f}  \,,
\qquad \qquad ~
C^{(2)}_{t\psi} ~=~ -{Qa\cos^2\theta\over Q+f}  \,,  \label{el}  \\
C^{(2)}_{y\phi}&=&-{Qa\sin^2\theta\over Q+f} \,, 
\qquad\quad 
C^{(2)}_{\phi\psi} ~=~ Q\cos^2\theta+{Qa^2\sin^2\theta\cos^2\theta\over Q+f}   \nnm 
\eea
where
\be
a={Q\over R_y}, \qquad f=r^2+a^2\cos^2\theta, \qquad h=1+{Q\over f} \,.
\ee
This configuration is a solution to the 10D equations of motion
\bea
R_{MN}&=&{1\over 4}F^{(3)}_{MPQ}F^{(3)}_N{}^{PQ}  \,,  \cr
F^{(3)}_{MNP}{}^{;P}&=&0 \,, \qquad\qquad   F^{(3)} ~=~ dC^{(2)} \,, \qquad \label{eq:eom1}
\eea
with $F^{(3)}$ being self-dual in 6D.

In this section we work in the regime of parameters
\be \label{eq:ring_limit}
\e ~~\equiv~~ \frac{R_y}{\sqrt{Q}} ~~\ll~~ 1 \, \qquad \Ri \qquad 
R_y ~~\ll \sqrt{Q} ~~\ll~~ a \,.
\ee
In this limit the background has the structure of a thin ring of radius $a$ and thickness $\sqrt{Q}$, as noted in~\cite{Lunin:2001fv}. The center of the ring is the locus $r = 0 $, $\theta = \pi / 2 $, which is a circle extended in the $\phi $ direction.

In the previous subsection, by considering dualities, we found that the configuration which corresponds to the fundamental string solution of Section~\ref{sec:string_soln} should have local dipole charges which oscillate from D1-D5 to NS1-NS5 around a thin ring. This motivates a guess for the fields sourced by such an object: 
start from \eq{el}, in the limit of a large ring \eq{eq:ring_limit}, and define new fields
\bea \label{eq:ansatz}
C^{(2)}{}' &=& C^{(2)} \cos \left[ k \left( \phi-\tfrac{t}{a} \right) \right] , \qquad
B^{(2)}{}' ~=~ C^{(2)} \sin \left[ k \left( \phi-\tfrac{t}{a} \right) \right] . \quad
\eea
If we replaced the $t$, $\phi$ dependence by a constant parameter, this would simply be an S-duality. Such an S-duality would not produce any dilaton $\varphi$ or R-R fields $C^{(0)}$ or $C^{(4)}$. The ansatz \eq{eq:ansatz} is too simple by itself to yield an exact solution, and we will not attempt to construct this solution here. Nevertheless, the duality argument suggests that such a solution exists, and we shall next see that, in the near-ring limit, these fields reduce to an exact supergravity solution with $\varphi$, $C^{(0)}$ and $C^{(4)}$ set to zero.

For the large ring, we expect the time-dependence in \eq{eq:ansatz} to give rise to non-zero fields $\varphi$, $C^{(0)}$ and $C^{(4)}$, as well as corrections to the metric. In particular, we expect this to lead to radiation in $\varphi$ and $C^{(0)}$. We will estimate the contribution to such radiation coming from the starting ansatz \eq{eq:ansatz} below, although we emphasize that we will not perform a systematic analysis.

\subsection{Exact oscillating supertube solution} \label{sec:nrlimit}

We now take a near-ring limit, in which we will obtain an exact solution for the oscillating supertube. The limit consists of `zooming in' very close to the ring, so that the coordinate along the ring becomes an infinite straight line to leading order. 

We first change to a more suitable coordinate system. Following~\cite{Giusto:2006zi}, we transform from $(r,\theta,\psi,\phi)$ to $(\rho,\vartheta,\varphi,\xi)$ with the coordinate transformation\footnote{To get to the coordinate naming conventions of \cite{Giusto:2006zi} one should relabel $(r,\theta,\psi,\phi) \to (\bar{r},\bar{\theta},\bar{\psi},\bar{\phi})$ and $(\rho,\vartheta,\varphi,\xi) \to (r,\theta,\phi,z)$.}
\bea
r^2 &=& {a^2 \rho (1-\cos\vartheta)\over a+\rho\cos\vartheta}\,,\quad 
\sin^2\theta ~=~ {a-\rho\over a+\rho\cos\vartheta} \,,\quad
\psi ~=~ \varphi \,, \quad \phi ~=~ {\xi\over a} \,. \qquad
\label{nrchange}
\eea
The coordinate $\rho$ measures distance from the ring. Near the ring, $\xi$ measures distance along the ring, and $r,\vartheta,\varphi$ are polar coordinates in the transverse three non-compact directions. $\vartheta$ is the polar angle with $\vartheta=0$ pointing towards the center of the ring.
Under this change of coordinates, we have 
\bea
\cos^2\theta ~=~ {\rho(1+\cos\vartheta) \over a+\rho\cos\vartheta} \,,
\qquad 
f &=& \frac{2 a^2 \rho}{a+\rho \cos \vartheta} \,.
\eea
We introduce $q$ for the charge density along the ring,
\be
q = \frac{Q}{2a}\,.
\ee
To take the near-ring limit we consider 
\be
\rho ~~\ll~~ a \,.
\ee
In this regime we have
\bea
f ~\simeq~ 2 a \rho \,, \qquad
r^2 &~\simeq~& a \rho (1-\cos\vartheta) \,.
\eea
The ansatz \eq{eq:ansatz} on the original background \eq{el} then reduces to the following fields:
\bea
ds^2&=&-H^{-1}\,\Bigl(dt+{q\over \rho} \, d\xi \Bigr)^2+H\,d\xi^2+ds^2_{TN} \,, \nonumber\\
ds^2_{TN}&=&H^{-1} \Bigl(dy- q(1+\cos\vartheta) \, d\varphi \Bigr)^2 
+ H\,\big(d\rho^2+\rho^2d\vartheta^2+\rho^2\sin^2\vartheta d\varphi^2 \big) \,, \nonumber\\
C^{(2)} &=& \cos \left[ \frac{k}{a}(\xi-t) \right] \left( H^{-1} \,{q\over \rho}\,dy\wedge (dt-d\xi)+H^{-1}  q(1+\cos\vartheta)\,d\varphi\wedge (dt-d\xi)
\right) \qquad \label{nr} \\
B^{(2)} &=& \sin \left[ \frac{k}{a}(\xi-t) \right] \left( H^{-1} \,{q\over \rho}\,dy\wedge (dt-d\xi)+H^{-1}  q(1+\cos\vartheta)\,d\varphi\wedge (dt-d\xi)
\right) \nonumber
\eea
where
\be
H=1+{q\over \rho} \,.
\ee
The fields $\varphi$, $C^{(0)}$ and $C^{(4)}$ may be consistently set to zero, so the type IIB supergravity field equations truncate to:
\bea
R_{MN}&=& T_{MN} \,, \qquad ~ T_{MN} ~=~ {1\over 4}F^{(3)}_{MPQ}F^{(3)}_N{}^{PQ}+{1\over 4}H^{(3)}_{MPQ}H^{(3)}_N{}^{PQ} \,, \cr 
F^{(3)}_{MNP}{}^{;P}&=&0 \,, \qquad\qquad   F^{(3)} ~=~ dC^{(2)} \,,  \label{eq:eom2} \\
H^{(3)}_{MNP}{}^{;P}&=&0 \,, \qquad\qquad   H^{(3)} ~=~ dB^{(2)} \,. \nonumber
\eea
By an explicit check, we find that the above fields are an exact solution to the equations of motion \eq{eq:eom2}. The resulting $F^{(3)}$ and $H^{(3)}$ are self-dual in 6D.

We interpret this solution as evidence that these oscillating supertubes exist as objects in string theory. This solution also suggests that the circular oscillating supertube ansatz \eq{eq:ansatz} captures at least some basic features of the desired circular oscillating supertube solution.

\subsection{Radiation from the ring configuration} \label{sec:rad}

The bound states discussed above are non-BPS; as is the case for all non-BPS microstates, they are expected to decay. Of course, atypical states may have shorter or longer lifetimes than typical states. It is interesting to investigate the physics of radiation from the above ring configuration.

As mentioned above, we expect that in the ring configuration, the time-dependence in the ansatz \eq{eq:ansatz} gives rise to radiation in $\varphi$ and $C^{(0)}$. We now make a simple estimate for the contribution to such radiation arising directly from the ansatz \eq{eq:ansatz}. We emphasize that we are not making a systematic analysis; a full investigation would involve an actual construction of the solutions, perhaps in a perturbative scheme. Nevertheless, let us make some preliminary remarks. To a first approximation, the dilaton and axion have almost identical behavior, and for concreteness we focus on the dilaton. We ignore numerical factors in the following and the details are presented in Appendix \ref{sec:app}. 

We remind the reader of the  small parameter  $\e$:
\be \label{eq:ring_limit2}
\e ~~\equiv~~ \frac{R_y}{\sqrt{Q}}~~=~~\frac{\sqrt{Q}}{a} ~~\ll~~ 1 \,.
\ee
In our setup, we have five large dimensions. We estimate the contribution to the dilaton from the fields in the starting ansatz \eq{eq:ansatz}. Such an estimate may be obtained by analyzing the  five-dimensional equation 
\bea
\square_5 \varphi ~\simeq~ \frac{1}{2}\left[ \big( F^{(3)} \big)^2 - \big( H^{(3)} \big)^2 \right]& \sim & \e^4k^2 \left[ \frac{a^2}{(r^2+a^2)(Q+f)} \cot^2 \theta \right] \cos \left [ 2k \left (\phi- \tfrac{t}{a} \right) \right]\qquad
\eea
where $\square_5 $ is the five dimensional d'Alembertian.
We focus on the parameters $\e$ and $a$. In Appendix \ref{sec:app}, we obtain the following estimate for the dilaton field:
\bea
\varphi(t,\mathbf{x}) & \sim & \e^4\, \frac{a^{\frac{3}{2}}}{r^{\frac{3}{2}}} 
%\int d^4\tilde{x}' 
%\frac{ \cot^2 \theta }{(\tilde{r}^2+1)(\tilde{r}^2+\cos^2\t\theta)} k^{\frac{5}{2}}
%\cos \left [ 2k \left (\tilde{t}-|\tilde{\mathbf{x}}-\tilde{\mathbf{x}}'|-\tilde{\phi}' \right) +\frac{\pi}{4}\right]
\int d^4\tilde{x}'
\frac{\cot^2\tilde{\th}'}{\left(\tilde{r}'^2+1\right)(\tilde{r}'^2+\cos^2\t\theta')} k^{\frac{5}{2}}\cos\left[2k\left(\tilde{t}-|\tilde{\mathbf{x}}-\tilde{\mathbf{x}}'|-\tilde{\phi}'\right)+\frac{\pi}{4}\right]
\qquad\label{eq:phisol}\\ 
&\equiv& \e^4 \,\frac{a^{\frac{3}{2}}}{r^{\frac{3}{2}}} \, I\left(\tilde{t}, \tilde{\mathbf{x}};k\right)\,,
\eea
where $\mathbf{x}$ is a vector in $\mathbb{R}^4$ and where variables with a tilde are dimensionless:
\bea
\tilde{t} ~=~ \frac{t}{a}\,, \quad\tilde{r} ~=~ \frac{r}{a}\,, \quad\tilde{\theta}~=~\theta\,, \quad\tilde{\phi}~=~\phi\,, \quad \mathrm{etc}\,.
\eea
In equation \eq{eq:phisol} we have extracted the parametric dependence on $\e$ and $a$, but not $k$. The $k$-dependence is sensitive to the details of the source near the core, which we have not constructed, so it is not possible to extract a reliable estimate at this level.

%While the $k$-dependence is left implicit above, we note that the oscillatory nature of the integrand is likely to lead to a behaviour such that larger $k$ leads to less radiation, despite the $\sqrt{k}$ in the numerator. This is to be expected, as large $k$ is related to the straight ring limit.

At infinity, to a good approximation we have flat Minkowski space, so that the energy flux is given by
\bea
T_{tr} &\simeq& \d_t \varphi \d_r \varphi
\eea
and the leading term arises when the derivatives fall on $I(\tilde{t}, \tilde{\mathbf{x}};k)$. Converting the derivatives to dimensionless variables, we find
\bea
T_{tr}&\sim&\e^8\frac{a}{r^3}\, \d_{\tilde{t}}I\left(\tilde{t}, \tilde{\mathbf{x}};k
\right) \, \d_{\tilde{r}} I\left(\tilde{t}, \tilde{\mathbf{x}};k
\right) \,.
\eea
Integrating over the $S^3$ at infinity, the  time-averaged power radiated to infinity is then\footnote{To restore dimensions, the denominator should receive a factor of the five-dimensional Newton constant.}:
\bea
\left\langle\frac{d E}{dt}\right\rangle
&=&\left\langle\lim_{r\to \infty}\int d\Omega_{(3)}r^3
 T_{tr}
\right\rangle\\
&\sim&\e^8\,a\,\left\langle\lim_{r\to \infty}
%\left[
\int d\Omega_{(3)}\, 
\d_{\tilde{t}}I\left(\tilde{t}, \tilde{\mathbf{x}};k
\right) \d_{\tilde{r}}I\left(\tilde{t}, \tilde{\mathbf{x}};k
\right) 
\right\rangle
%\right]
\,.%, \qquad\quad \e ~=~  ~\ll~ 1 \,.
\eea

Let us make some comments on this estimate. Firstly, if we set $k$ to be some fixed number of order unity, we note that the radiation rate is parametrically small, controlled by the parameter $\e$. If we re-express $\e$ in terms of $R_y$ and $a$, we see that the radiation rate decreases with increasing $a$. For example, if we set $R_y$ (and all other compact directions) to be of order the Planck length, for $k$ of order unity we find
\bea
\left\langle\frac{d E}{dt}\right\rangle
&\sim&\frac{l_p}{a^3}\,.
\eea

If this estimate accurately represents the radiation rate from these configurations, it suggests they may be very long-lived. However, there are many uncertainties which should be addressed before one can draw such a conclusion.
Firstly, we cannot have full confidence in the estimate itself; a priori, it is possible that a full construction of the solution will lead to a greater or lower radiation rate. 
Secondly, as noted above, at this level it is not possible to estimate the $k$-dependence of the radiation rate.
Finally, we cannot have full confidence that the radiation in the dilaton and axion is the dominant decay channel. Clearly a complete systematic analysis is required. Once a full investigation has been carried out, it would be interesting to 
investigate further the suggestion that such radiation may be related to superradiance~\cite{BlancoPillado:2007iz}. 

In addition, one is naturally led to wonder about more complicated configurations, involving arbitrary profile shapes, and larger $k$. It is a very interesting question to ask whether the lifetime of generic configurations matches the lifetime of the corresponding black hole solutions. The answer will depend on many factors, including the typical local extrinsic curvature radius of the generic profile, the $k$ dependence, cancellations between different segments of the configuration, and the redshift between the source and the asymptotics. In the present configuration the redshift is negligible, but for more complicated profiles we expect this to be important. Such questions are beyond the scope of this paper, and this is an interesting direction for future study.

%\newpage
\section{Neutral non-BPS states from JMaRT solutions}

In this section we conjecture the existence of an oscillating supertube solution based on the two-charge JMaRT solutions. We first review these solutions and give an overview of our conjecture.

\subsection{Overview}

Most microstate constructions to date have focused on supersymmetric solutions, such as the one in Section \ref{sec:MM_ring} above. 
Two-charge BPS D1-D5 microstates can all be constructed~\cite{\lmfour,\lmm,\kst}, and studied in a 1+1 dimensional CFT living along the common direction $y$ shared by the D1 and D5 branes. We focus on axisymmetric supergravity solutions. In the CFT we can consider the operation of spectral flow \cite{Schwimmer:1986mf}, which maps states of the theory to new states.  In \cite{\lunin,\gmsone,*\gmstwo} spectral flow on the left movers was considered. The CFT state acquires a momentum and an additional energy which are equal. Thus we obtain BPS states with momentum charge in the $y$ direction.  The corresponding geometries were constructed, and found to be regular supergravity solutions.

In \cite{Jejjala:2005yu}, microstate geometries were found which are non-BPS.
These solutions correspond to CFT states where spectral flow is performed on both the left and the right movers. Suppose we add $n_p$ units of momentum to the left-moving sector and $\bar{n}_p$ units of momentum to the right-moving sector. The net momentum charge of the CFT state is $n_p-\bar{n}_p$, while the added energy is $n_p+\bar{n}_p$. Thus the state may have more energy than charge, making it non-BPS. 

Our goal is to see if we can make states that are not just non-BPS, but {\it neutral}. We proceed in the following steps:

\b

(i) We consider an equal amount of spectral flow $m $ on both left and right sectors, so that  we add no net momentum charge to the state. For convenience we set the D1 and D5 charges equal: $Q_1=Q_5=Q$.

\b

(ii) The length of the $y$ circle, $R_y$, must be taken large ($R_y\gg \sqrt{Q}$) to get a limit where we have a large AdS region and thus a well-defined dual CFT. But the geometries of \cite{Jejjala:2005yu} allow us to choose any value for $R_y$, and we take $R_y$ {\it small},
\bea \label{eq:smallry}
R_y & \ll & \sqrt{Q} \,.
\eea
This is because if we wish to describe solutions that have the properties of black holes in the non-compact directions, we must take the compact directions to be small compared to the length scales involved in the non-compact part of the black hole geometry.

\b

(iii) We consider the regime of parameters for which the D1 and D5 charges are much smaller than the ADM mass of the solution. This requires us to consider large values of the spectral flow parameter $m$,
\bea
m &\gg& 1 \,,
\eea
and we shall discuss precisely how large in due course.

\b

(iv) We then observe that these small D1, D5 charges influence the geometry only in a region which is shaped like a thin ring. The local structure is therefore similar to that of the ring solution which we studied in Section \ref{sec:D1D5ring}. In the previous case the spacetime was flat away from the ring, while in our present case the spacetime has a nonzero curvature away from the ring. But the length scale of this curvature is large compared to the thickness of the ring, so we have a ring that is sitting in gently curved space. We then conjecture that, as with the two-charge ring, we can allow the charges to oscillate between D1-D5 and NS1-NS5 along the ring direction, in such a way that the overall charge of the ring is zero. 

\b

(v) If the above conjecture is correct, then this would be a new solution that has no conserved gauge charges. It has large angular momentum, generated by the spectral flows we have taken. In fact, the 5D Myers-Perry solution with the same mass and angular momentum is slightly over rotating, and so has a naked ring singularity. The JMaRT solutions are however all smooth and singularity-free, so we can think of the Myers-Perry singularity as being resolved by the small D1 and D5 charges. The parameters may be chosen to make the over rotation arbitrarily small, so this would be a new solution which has approximately the quantum numbers of the extremal 5D singly-spinning Myers-Perry black hole. We discuss obtaining solutions with less angular momentum in the Discussion section.
%

%\newpage 
\subsection{Two-charge non-BPS horizonless solutions}

We start with the smooth horizonless solutions studied in~\cite{Jejjala:2005yu}. These solutions are special cases of the general rotating three-charge solutions obtained in \cite{Cvetic:1996xz}, which include black hole solutions as well as smooth solitons.

We focus on the subset of solutions with no net momentum charge, and for simplicity we set the D1 and D5 charges to be equal, $Q_1 = Q_5 = Q$. With this choice of parameters, and using the shorthand notation $c = \cosh \delta$, $s = \sinh \delta$, the metric simplifies to\footnote{Starting from the general solution in Eq.~(2.1) of~~\cite{Jejjala:2005yu}, the above solution is obtained by setting $a_1=\delta_p=0$ and $\delta_1=\delta_5=\delta$, giving $Q_1 = Q_5 = Q$ etc.}
\begin{eqnarray}  \label{2chmetric}
ds^2&=&
\tilde{H}^{-1}
\left[- \left(f-M \right) \left(dt - \frac{a_2 M}{(f-M)}\sin^2\theta \,c^2 \, d\phi \right)^2 
 \right.\\
&& \qquad\qquad\qquad\quad \left. 
{} + f \left( dy +  \frac{a_2 M}{f} \cos^2\theta \, s^2 \, d\psi \right)^2 \right] \nonumber \\
&& {} + \tilde{H} \left(\frac{dr^2}{r^2+a_2^2-M}  
+d\theta^2+\frac{r^2\cos^2\theta}{f}d\psi^2
+\frac{(r^2+a_2^2-M)\sin^2\theta}{f-M}d\phi^2\right) \quad \nonumber
\end{eqnarray}
where
\bea 
\tilde{H} ~=~ f+M\sinh^2\delta \,, \qquad f ~=~ r^2+a_2^2\cos^2\theta \,.
\eea
The RR 2-form simplifies to
\begin{eqnarray} 
C_2 &=& - \frac{Q}{\tilde H} dt \wedge dy + \frac{Q \, a_2}{\tilde H} \left( \cos^2 \theta \, dt  \wedge d\psi
 +  \sin^2 \th \, dy  \wedge d \phi \right) \quad
  \nonumber \\
&&  -  \frac{Q}{\tilde H} (r^2 + a_2^2 + M  s^2) \cos^2 \theta \, d\psi \wedge d\phi \,, \nonumber
\end{eqnarray}
where the five-dimensional D1 and D5 charge $Q$ is given by
\be \label{eq:5DchargeQ}
Q ~=~ M \sinh \delta \cosh \delta \,.
\ee
For smooth horizonless solutions we have the following constraints:
\bea \label{eq:2chperiod}
\frac{M s^2}{\sqrt{a_2^2-M}}  &=& R_y \,, \\
\label{eq:m}
\frac{a_2}{\sqrt{a_2^2-M}} &=& m ~\in~ \mathbb{Z} \,.
\eea
In units in which $4 G^{(5)}/\pi =1$, the Einstein frame ADM mass is
\bea \label{eq:5DchargeM}
M_{ADM} &=& M \cosh 2\delta \,. 
\eea

\subsection{Regime of parameters of interest}

In the above solution there are three free parameters, which we shall take to be $Q$, $R_y$, $m$. As discussed above, we take $R_y \ll \sqrt{Q}$ and we work far from the BPS limit, $2Q ~\ll~ M_{ADM}$. From \eq{eq:5DchargeQ}, \eq{eq:5DchargeM}, we see that this implies that $\delta \ll 1$, so we have two small parameters:  
\bea \label{eq:ep1r}
\frac{R_y}{\sqrt{Q}} ~=~ \e_1  &\ll& 1 \,, \\
\delta &\ll& 1 \,
\eea
and we also have the relation
\bea
Q~\simeq~ M_{ADM} \,\delta \,.
\eea
We are mainly thinking of the D1 and D5 charge $Q$ as a fixed regulator, which resolves the ring singularity of the corresponding 5D singly spinning Myers-Perry solution. As a result we think of the parameter $\epsilon_1 $ as some fixed small number, say $10^{-3}$.

On the other hand, we would like to have an arbitrary large hierarchy between the charge and the ADM mass, which we think of as macroscopic. Therefore we require $\delta $ to be arbitrarily small.
Combining equations \eq{eq:5DchargeQ}--\eq{eq:m}, one may show that
\bea \label{eq:de}
 \frac{\e_1^2}{(m^2-1)} &=& \frac{\sinh^3 \delta}{\cosh \delta}  ~\;\simeq~\; \delta^3
\eea
where in the second step we have taken the small $\delta $ approximation. To make $\delta $ arbitrarily small, we must therefore take $m$ to be correspondingly large, according to \eq{eq:de}.

In this regime of parameters,  the 5D solution resembles a thin ring inside a 5D singly-spinning Myers-Perry solution. 
In the limit $m \to \infty$, $\delta \to 0$, the 5D metric approaches the 5D extremal singly-spinning Myers-Perry solution.

\subsection{Conjecture}

We have seen that in a particular regime of parameters, the two-charge JMaRT solutions resemble a thin charged ring inside a Myers-Perry solution. We now conjecture that a similar construction to that of Section 3 yields a new charge-neutral solution.

In the present case, we have a thin charged ring inside a Myers-Perry solution, rather than a thin charged ring in flat space as in the previous section. As a result, we expect the solution, if it exists, to be more involved than that of the previous section\footnote{For example, it is not clear to us whether a useful analog of the coordinates used in Section \ref{sec:nrlimit} exists.}. A detailed investigation is beyond the scope of this paper, but by analogy with the previous example, we expect that a starting point should be a similar ansatz resembling a ``time-dependent S-duality'',
\bea
C^{(2)}{}' &=& C^{(2)} \cos \left[ k \left( \phi-\tfrac{t}{a_2} \right) \right] , \qquad
B^{(2)}{}' ~=~ C^{(2)} \sin \left[ k \left( \phi-\tfrac{t}{a_2} \right) \right] . \quad
\eea
If such a solution exists, this would be very interesting in its own right. As before, there are interesting questions regarding the stability of any such solution. The JMaRT backgrounds themselves have an ergoregion instability~\cite{Cardoso:2005gj,
Chowdhury:2007jx,*Chowdhury:2008bd,*Chowdhury:2008uj}, and we are introducing time-dependent gauge fields.

Nevertheless, we are interested in the possibility that this type of construction may play a role in finding solutions which carry less angular momentum, and which would thus be candidates to be identified as microstates of macroscopic neutral black holes. We comment further on this in the Discussion.

\section{Discussion}

In this paper we have studied bound states of string theory which are charge neutral; the configurations we studied have dipole charges but no conserved charges other than mass and angular momentum. We focused on states with large angular momentum, which provides a simplification in the construction of microstates.

We started with a fundamental string configuration with the structure of a ring, which has local dipole winding and momentum charges. The charges rotate around the ring in such a way that there is no overall conserved winding or momentum charge. By considering a sequence of S and T dualities, we inferred the existence of an `oscillating supertube', whose local dipole charges oscillate as a null wave along the supertube profile (in our example, from D1-D5 to NS1-NS5). 
We found an exact supergravity solution in the near-ring limit, in which the supertube profile becomes an infinite straight line. The solution is everywhere smooth. We regard this as evidence that this oscillating supertube exists as an object in string theory. We also made a simple estimate for the radiation rate from the ring configuration. 

We then used this result to conjecture that a similar construction may be applied to a set of two-charge JMaRT solutions. In a particular regime of parameters, this family of solutions has the structure of a thin charged ring inside a 5D singly-spinning Myers-Perry solution. 
The Myers-Perry solutions include both black holes with classical horizons and solutions which have naked singularities at the classical level. In the 5D singly-spinning case, when the angular momentum is below the extremality bound there is a classical horizon;  when the angular momentum is extremal or over-rotating, the solutions have naked ring singularities. The JMaRT solutions have quantum numbers in the over-rotating regime, but the over-rotation can be made arbitrarily small, approaching the extremal value.

Using the intuition from the two-charge ring, we conjectured that we could alter these two-charge JMaRT solutions to make the charge change with position along the ring, setting the overall charge to zero. This would then be a regular, horizonless, neutral bound state with quantum numbers arbitrarily close to those of the extremal 5D singly-spinning Myers-Perry solution. 

Our work offers many opportunities for future research. Most obvious are the explicit construction of circular oscillating supertube solutions, and a detailed study of their decay modes. Most interesting to us however, is the question of whether this line of thinking can be taken further. We would of course like to understand 
microstates of black holes with macroscopic classical horizons. 
In the present setting of 5D singly-spinning Myers-Perry black holes, this means understanding microstates with angular momentum below the extremality bound. It would also be interesting to see if it is possible to construct similar solutions corresponding to microstates of other neutral black objects, such as black rings (see e.g.~\cite{Emparan:2006mm}) and helical black rings~\cite{Emparan:2009vd}.

One way that such microstates might be realized is as follows. The thin ring carrying charge in the above Myers-Perry-like geometry is similar, locally, to the supertube carrying charge in the two-charge case which we studied in Section~\ref{sec:D1D5ring}. In the latter case we can think of the supertube as `sourcing' the rest of the geometry: one can  choose any shape for the supertube, and a corresponding geometry will be generated. Different supertube profiles give rise to different geometries, and in particular we can choose profiles that have no overall angular momentum.

It would be interesting to know, in the case of the Myers-Perry-like microstate geometries we have conjectured, whether a similar procedure could be followed. First, we can ask if the thin tube carrying charge could be considered as `sourcing' the rest of the geometry. If this is true, then it would appear likely that we can consider any shape we wish for this tube, including profiles that have no overall angular momentum. In that case the geometry sourced by the tube would have no angular momentum either. Since the configuration was constructed to have no conserved gauge charges, the geometry would have neither charge nor angular momentum; thus we would obtain a bound state with the quantum numbers of the Schwarzschild black hole.

\section*{Acknowledgements}

We thank Stefano Giusto, Amanda Peet, Simon Ross, Rodolfo Russo, 
and especially Iosif Bena and Nick Warner for fruitful discussions. 
We thank the anonymous referee for comments which led to improvements in the paper.
This work was supported in part by DOE grant DE-FG02-91ER-40690.

%\newpage

\begin{appendix}

\section{Radiation estimate}\label{sec:app}

We expect the starting ansatz \eq{eq:ansatz} to give rise to non-zero dilaton $\varphi $ and RR fields $C^{(0)}$ and $C^{(4)}$, as well as corrections to the metric. In particular, we expect radiation in $\varphi$ and $C^{(0)}$. Here we make a simple estimate for this radiation. 
For concreteness, we will mainly focus on the dilaton.

%We first linearize in $\e$ the relevant equations of motion. 
We start in 10D Einstein frame, in which the type IIB supergravity equations of motion are 
\bea
\sq \varphi &=& e^{2\varphi}\del_M C^{(0)}\del^M C^{(0)}
+\frac{1}{12} e^{\varphi} \t F^{(3)}_{MNP}(\t F^{(3)})^{MNP}
%\nonumber\\
%&&\hspace{3cm}
-\frac{1}{12} e^{-\varphi} H^{(3)}_{MNP}(H^{(3)})^{MNP}\,,\qquad\\
\sq C^{(0)} &=& -2\del_M\varphi\del^M C^{(0)}-\frac{1}{6} e^{-\varphi} \t F^{(3)}_{MNP}(H^{(3)})^{MNP}
\,,  %~~\\
\eea
\vspace{-8mm}
\bea
d\star\left(e^\varphi \t F^{(3)}\right)&=& \t F^{(5)}\wedge H^{(3)} \,,\\
d\star\left(e^{-\varphi}H^{(3)}-e^\varphi C^{(0)}\t F^{(3)}\right)&=&-\t F^{(5)}\wedge F^{(3)} \,,\\
d\star\t F^{(5)}&=& H^{(3)}\wedge \t F^{(3)} \,, \qquad\qquad \t F^{(5)}~=~\star\,\t F^{(5)}\,,\qquad\qquad
\eea
along with the Einstein equation, where 
\bea
&&{}\t F^{(3)} ~=~ F^{(3)} - C^{(0)}H ^{(3)} \,,\qquad %\\
\t F^{(5)} ~=~ F^{(5)} - C^{(2)}\wedge H ^{(3)} \,,\\
&&{}H^{(3)}~=~dB^{(2)}\,, \qquad F^{(3)}~=~dC^{(2)}\,, \qquad F^{(5)}~=~dC^{(4)}\,.
\eea
The starting ansatz \eq{eq:ansatz} gives contributions as follows (using $(F^{(3)}) ^ 2=\tfrac{1}{6}F^{(3)}_{MNP} (F^{(3)})^{MNP}$ and similarly for $H^{(3)}$)
\bea
(F^{(3)})^2 &\sim& \e^4 k^2 \left[ \frac{a^2}{(r^2+a^2)(Q+f)} \cot^2 \theta \right] \sin^2 \left [ k \left (\phi- \tfrac{t}{a} \right) \right]  \,,\\
(H^{(3)})^2 &\sim& \e^4 k^2 \left[ \frac{a^2}{(r^2+a^2)(Q+f)} \cot^2 \theta \right] \cos^2 \left [ k \left (\phi- \tfrac{t}{a} \right) \right]  \,,\\
F^{(3)}_{MNP}(H^{(3)})^{MNP} &\sim& \e^4 k^2 \left[ \frac{a^2}{(r^2+a^2)(Q+f)} \cot^2 \theta \right] \sin \left [ 2k \left (\phi- \tfrac{t}{a} \right) \right]  \,,\\
C^{(2)} \wedge B^{(2)} &\sim&\e^4
\frac{a^4\cos ^2\theta}{\left(Q+f\right)}
\sin \left[2 k \left(\phi -\tfrac{t}{a}\right)\right] dt\wedge dy \wedge d\psi \wedge d\phi\,, 
%\\
%
%C^{(2)} \wedge H &\sim&
%-\frac{Q^2 r \cos ^2\theta}{\left(Q+f\right)^2}
%\sin \left[2 k \left(\phi -\tfrac{t}{a}\right)\right] dt\wedge dy\wedge dr \wedge d\psi \wedge d\phi \\
%%
%&&{}\hspace{-12mm}-\frac{Q^2 \sin \theta \cos \theta  \left(Q+r^2\right) }{\left(Q+f\right)^2}
%\sin \left[2 k \left(\phi -\tfrac{t}{a}\right)\right] dt\wedge dy\wedge d\th \wedge d\psi \wedge d\phi \\
%\star\left(C^{(2)} \wedge H \right) &\sim&
%\frac{Q^2 (Q+r^2)}{r(r^2+a^2)\left(Q+f\right)^2}
%\sin \left[2 k \left(\phi -\tfrac{t}{a}\right)\right] dr \wedge dT^4 \\
%%
%&&{}
%%\hspace{-12mm}
%-\frac{Q^2 \cot \theta }{\left(Q+f\right)^2}
%\sin \left[2 k \left(\phi -\tfrac{t}{a}\right)\right] d\th \wedge dT^4
\eea
where as before $f=r^2+a^2\cos^2\theta $.

The dilaton equation of motion, to leading order in $\epsilon$, becomes 
\bea
\square \varphi & \simeq & \frac{1}{2}\left[ \big( F^{(3)} \big)^2 - \big( H^{(3)} \big)^2 \right]
~\sim~
 \e^4 k^2 \left[ \frac{a^2}{(r^2+a^2)(Q+f)} \cot^2 \theta \right] \cos \left [ 2k \left (\phi- \tfrac{t}{a} \right) \right].\qquad\quad\,.\label{eq}
\eea
Similarly, the axion equation of motion, to leading order in $\epsilon$, becomes
\bea
\square C ^ {(0)}& \simeq & -\frac{1}{6}F^{(3)}_{MNP}(H^{(3)})^{MNP} %\,\\
~\sim~
 \e^4 k^2 \left[ \frac{a^2}{(r^2+a^2)(Q+f)} \cot^2 \theta \right] \sin \left [ 2k \left (\phi- \tfrac{t}{a} \right) \right].\qquad\quad
\eea
In passing, we note that the non-zero $C^{(2)} \wedge B^{(2)}$ will contribute to the $C^{(4)}$ field. 

We now focus on the dilaton and make a simple estimate for the field in the radiation zone. As one can see from the above expressions, the $C ^ {(0)}$ calculation is almost identical. 

In our setup, we have five large dimensions. Let us write the five-dimensional version of \eq{eq} as  ($\square_5 $ is the five dimensional d'Alembertian)
\bea \label{eq:5dwave}
\square_5 \varphi & = & S (t,\mathbf {x}) \,,
\eea
where the source $ S (t,\mathbf {x}) $ is
\bea \label{eq:source}
S (t,\mathbf {x}) & = & \e^4k^2 \left[ \frac{a^2}{(r^2+a^2)(Q+f)} \cot^2 \theta \right] \cos \left [ 2k \left (\phi- \tfrac{t}{a} \right) \right] \,.
\eea

%\subsection{Five Dimensions}
Rather than using the explicit form of the source immediately, let us start with a general source $ S (t,\mathbf {x}) $. The retarded Green's function for the wave equation in $4+1$ dimensions, ignoring numerical factors, is (see e.g.~\cite{Soodak:1993,Hassani:1999,Cardoso:2002pa})
\bea \label{eq:gret}
G^{\mathrm{ret}}(t,\mathbf{x}) &\sim& 
\frac{\delta(t-r)}{r\sqrt {t ^ 2 - r ^ 2}} - \frac{\Theta(t-r)}{(t ^ 2 - r ^ 2)^{3/2}}\,,
\eea
where $r = |\mathbf {x} | $. For a general source $ S (t,\mathbf {x}) $ we have
\bea \label{eq:phi1}
\varphi (t,\mathbf {x}) & = &\int d^4x' \!\int\limits_{-\infty} ^ {\infty} dt' \,S(t',\mathbf {x'}) \,G^{\mathrm{ret}}(t -t',\mathbf{x -x'}) \,.
\eea
Let us introduce $\scripty{r} =|\mathbf{x} -\mathbf{x'} | $ and write
\bea \label{eq:deriv}
\frac{1}{\left[ (t-t') ^ 2 - \scripty{r} ^ 2\right]^{3/2}}&=& 
 \frac {1} {t -t'} \frac {\d} {\d t'} \frac{1}{\sqrt {(t-t') ^ 2 - \scripty{r} ^ 2}} \,.
\eea
Then \eq{eq:phi1} becomes
\bea \label{eq:phi2}
\varphi (t,\mathbf {x}) &\!\!\sim \!\!\!&\int \!d^4x' \!\!\int\limits_{-\infty} ^ {\infty} \!\!dt'\, S(t',\mathbf {x'}) 
\left[ \frac{\delta((t-t') - \scripty{r})}{\scripty{r}\sqrt {(t-t') ^ 2 - \scripty{r} ^ 2}} -
\Theta((t-t') - \scripty{r})
 \frac {1} {t -t'} \frac {\d} {\d t'} \frac{1}{\sqrt {(t-t') ^ 2 - \scripty{r} ^ 2}}
\right] .\nn
\eea
We integrate the second term by parts in $t'$, whereupon one of the resulting terms cancels the first term. We thus obtain
\bea
\varphi (t,\mathbf {x})& \sim &\int \!d^4x' \!\!\int\limits_{-\infty} ^ {t-\scripty{r}}\! dt' \,
\left [ \frac {S ^ {(1, 0)} (t',\mathbf {x'})} {t -t'} -\frac {S (t',\mathbf {x'})} {(t -t') ^ 2} \right] \frac{1}{\sqrt{(t - t')^2-\scripty{r}^2}}  \, \label{eq:s1}
\eea
where $S ^ {(1, 0)} (t',\mathbf {x'})$ denotes the derivative of $S(t',\mathbf {x'})$ with respect to its first argument. 

Now let us consider the source of interest, given in Eq.~\eq{eq:source}. We also focus on the first term in \eq{eq:s1}. The resulting $t'$ integral is
\bea
I_{t'} &=&\int \limits_{-\infty}^{t- \scripty{r}} dt'\,
\frac{k} {a} \,\frac {\sin\left[ {2k \left (\phi'- \tfrac {t'} {a} \right)}\right]} {
\sqrt {(t- \scripty{r}) -t'} \,\,(t-t')\,\sqrt {(t + \scripty {r}) -t'}}
\,.
\eea
Since we are ignoring overall numerical factors, in the second and third factors of the denominator we may make the approximation
\bea
t- t'\simeq \scripty{r} \,.
\eea
We have verified this numerically. The integral then becomes
\bea
%&&
I_{t'}  ~\sim~\frac{k}{a} \frac {1} {\scripty{r}^{3/2}}\int\limits_{-\infty}^{t-\scripty{r}}
dt'\,\frac {\sin \left[{2k \left (\phi'- \tfrac {t'} {a} \right)}\right]} {
\sqrt {(t- \scripty{r}) -t'} }
%\\
&\sim& \sqrt{\frac{k}{a}}\frac{1}{\scripty{r}^{\frac{3}{2}}}\cos\left[\frac{2k}{a}\left(t-\scripty{r}-a\phi '\right)+\frac{\pi}{4}\right]\,.\quad
\eea
A similar analysis shows that the second term in \eq{eq:s1} is subleading for large $|\mathbf{x}|$, so we discard it.
For large $|\mathbf{x}|$ we then obtain
\bea
\varphi(t,\mathbf{x})&\sim&\e^4k^2\,\frac{a^{\frac{3}{2}}}{|\mathbf{x}|^{\frac{3}{2}}}\int d^4x'
\frac{\cot^2\th'}{\left(r'^2+a^2\right)(Q+f)} \sqrt{k}\cos\left[\frac{2k}{a}\left(t-|\mathbf{x}-\mathbf{x}'|-a\phi '\right)+\frac{\pi}{4}\right]\,.\qquad\quad
\eea
Let us estimate the dependence of the $\mathbf{x}'$ integral on the parameters $\e$ and $a$ (the $k$-dependence is sensitive to the details of the source near the core, which we have not constructed, so it is not possible to extract a reliable estimate at this level). 
The main scale in the problem is $a$, so we introduce the
dimensionless variables:
\bea
\tilde{t}' ~=~ \frac{t'}{a}\,, \quad\tilde{r}' ~=~ \frac{r'}{a}\,, \quad\tilde{\theta}'~=~\theta'\,, \quad\tilde{\phi}'~=~\phi'\,, \quad \mathrm{etc}\,.
\eea
and similarly for unprimed variables.
Let us also make the approximation
\bea
Q+f&=&Q+r'^2+a^2\cos^2\th'%\\
~=~a^2\left(\t r'^2+\cos^2\t\theta'+\e^2\right) ~\simeq~a^2\left(\t{r}'^2+\cos^2\t{\theta}'\right)
.
\eea
Then the $a$ dependence of the measure factor and denominator cancel, yielding
\bea
\varphi(t,\mathbf{x})&\sim&\e^4\,\frac{a^{\frac{3}{2}}}{|\mathbf{x}|^{\frac{3}{2}}}
\int d^4\tilde{x}'
\frac{\cot^2\tilde{\th}'}{\left(\tilde{r}'^2+1\right)(\tilde{r}'^2+\cos^2\t\theta')} k^{\frac{5}{2}}\cos\left[2k\left(\tilde{t}-|\tilde{\mathbf{x}}-\tilde{\mathbf{x}}'|-\tilde{\phi}'\right)+\frac{\pi}{4}\right].
\qquad\quad
\eea
%
%{eq:5dwave}

\end{appendix}

\newpage

\begin{adjustwidth}{-2.5mm}{-2.5mm}% adjust the L and R margins

\bibliographystyle{utphysM}      
%
% uses file "utphysM.bst", a modification of
% utphys.bst compatible with mciteplus package for combined citations.
%
\bibliography{microstates}       % uses file "microstates.bib"

\end{adjustwidth}

\end{document}